# Study on Thickness and Temperature Dependence of Thermoelectric Properties in SnS Nanofilms*

CHEN Liqi[1], WANG Ziyang[1], LI Donghao[1], WANG Jingye[1], ZHAO Ning[2], ZHOU Jun[2], ZHU Jie[1, *], TANG Dawei[1]

1. School of Energy and Power Engineering, Dalian University of Technology, Dalian 116000, China
2. School of Physics and Technology, Nanjing Normal University, Nanjing 210023, China

**Abstract:**

SnS as an environmentally friendly, cost-effective, and earth-abundant narrow-bandgap semiconductor material, has demonstrated significant application potential in the field of medium-temperature thermoelectric conversion. However, the thermoelectric performance of its bulk counterpart is inherently constrained by intrinsic point defects (e.g., vacancies) and the material's specific band structure. Low-dimensional engineering has emerged as a pivotal strategy for overcoming these limitations and enhancing thermoelectric performance. In this work, we systematically investigate the thermoelectric properties of SnS nanofilms with distinct thicknesses (82 nm, 199 nm, 616 nm, and 813 nm) across a temperature range of 300 – 600 K. Measurements were conducted using time-domain thermoreflectance (TDTR) and a dedicated thin-film thermoelectric parameter test system (ZEM-3). Our results confirm that low-dimensionalization effectively boosts the thermoelectric performance of SnS, with the thermoelectric figure of merit (*ZT*) displaying a pronounced dependence on both film thickness and temperature. All four SnS thin films exhibit thermoelectric performance that is markedly superior to that of bulk SnS. This enhancement is primarily attributed to the quantum confinement effect, energy filtering effect, and intensified phonon scattering, all of which are induced by the low-dimensional structural characteristics. This work provides not only experimental

---

evidence and theoretical insights for the performance optimization of SnS nanofilms but also establishes a foundational framework for the development of high-efficiency, eco-friendly medium-temperature thermoelectric materials, thereby holding significant scientific value and practical implications.

# 1 Introduction

In response to the escalating global energy crisis and environmental challenges, thermoelectric materials have emerged as a prominent sustainable energy technology due to their unique ability to directly convert waste heat into electricity[1,2]. As all-solid-state energy conversion media, these materials facilitate the direct interconversion of thermal and electrical energy through the directional migration of internal charge carriers[3]. Owing to their distinct conversion mechanisms, multifunctionality, and environmental compatibility, thermoelectric materials are widely employed in cutting-edge fields such as the automotive industry[4], space probes[4], solar thermal systems[5], and implantable or wearable devices[6]. However, the conversion efficiency and fabrication costs of existing thermoelectric materials remain critical barriers to their large-scale commercialization. Consequently, the search for novel thermoelectric materials that offer both low cost and high performance has become a focal point of current research[7-9]. The core metric for evaluating thermoelectric conversion efficiency is the dimensionless figure of merit ($ZT$)[10], which is expressed as

$$ZT = S^2\sigma T/\kappa, \tag{1}$$

Here, $S$ denotes the Seebeck coefficient, $\sigma$ represents the electrical conductivity, $T$ is the absolute temperature, and $\kappa$ indicates the thermal conductivity. The power factor (PF) is commonly used to comprehensively characterize the electrical transport properties of materials.

$$\mathrm{PF} = S^2\sigma. \tag{2}$$

To achieve efficient thermoelectric conversion, ideal thermoelectric materials should simultaneously possess a high Seebeck coefficient, high electrical conductivity, and low thermal conductivity to attain a higher $ZT$ value[11]. However, these parameters are often strongly coupled; for instance, high electrical conductivity is typically accompanied by high thermal conductivity, posing challenges for the optimization of thermoelectric materials[12].

Group IV-VI semiconductor materials, such as GeTe and PbTe, frequently exhibit high Seebeck coefficients and electrical conductivity within specific temperature ranges due to their appropriate band gaps and electronic structures. Consequently, they have long been a focal point of thermoelectric research[13]. Furthermore, carrier concentration[14] and band structure[15] in these compounds can be precisely tuned via doping or alloying, thereby further optimizing their electrical transport properties. Nevertheless, the low crustal abundance of elements like Te and Ge, along with the toxicity of Pb, restricts their large-scale commercial application. In recent years, SnSe has rapidly attracted significant academic attention due to its strong anharmonicity, multivalley band structure, and "three-dimensional charge, two-dimensional phonon" transport mechanism, which contribute to its superior thermoelectric performance[12]. As an analog of SnSe, SnS not only shares the same orthorhombic crystal structure and highly similar band characteristics but also features abundant sulfur reserves, environmental friendliness, and low cost, aligning better with the development direction of sustainable green energy materials[12]. Moreover, the inherently low thermal conductivity and high Seebeck coefficient of SnS confirm its substantial potential as a high-quality thermoelectric candidate[8]. Therefore, conducting multidimensional research on the thermoelectric properties of SnS is not only significant but also provides potential pathways for promoting practical applications in materials science and energy technology.

To further optimize thermoelectric performance, achieving material "low-dimensionality" through structural design has become one of the mainstream strategies in recent years[16]. Over the past few decades, research on low-dimensional thermoelectric materials has made substantial progress[17]. Regarding thermal transport, studies indicate that low-dimensional materials can introduce additional interface scattering through size effects, effectively reducing the phonon mean free path and thereby suppressing lattice thermal conductivity. In terms of electrical transport, the quantum confinement effect induced by low-dimensional structures leads to the discretization of electronic energy levels and distorts the electron density of states near the Fermi level[18]. This modification of the electronic structure facilitates a significant enhancement of the Seebeck coefficient while maintaining high electrical conductivity[9]. Due to this dual regulation of thermal and electrical properties, low-dimensional materials demonstrate greater potential for thermoelectric performance than bulk materials[18,19]. For thin-film materials, thermoelectric properties often exhibit significant thickness dependence[20]. Additionally, temperature, as a key factor influencing carrier transport, has a decisive impact on the figure of merit[21-23]. Therefore, systematically investigating the synergistic effects of thickness and temperature on the thermoelectric properties of SnS thin films holds important guiding value for developing efficient and stable thermoelectric conversion devices[24].

In this work, SnS thin films with four different thicknesses (82 nm, 199 nm, 616 nm, and 813 nm) were prepared using magnetron sputtering. The thermoelectric properties of these four SnS thin films were characterized in the temperature range of 300-600 K using the time-domain thermoreflectance (TDTR) method and a thin-film thermoelectric parameter measurement system (ZEM-3). TDTR provides accurate characterization of the thermal conductivity of nanoscale thin films, while ZEM-3 is used to comprehensively measure the Seebeck coefficient and electrical conductivity of the films. Simultaneously, scanning electron microscopy (SEM) and energy-dispersive X-ray spectroscopy (EDS) were employed to characterize the thickness and elemental composition. Based on these characterizations, the physical mechanisms by which thickness and temperature affect the Seebeck coefficient, electrical conductivity, thermal conductivity, and *ZT* value were analyzed. This study aims to elucidate the potential mechanisms by which low-dimensionality enhances the thermoelectric performance of SnS materials and to highlight their guiding significance for practical applications.

# 2 Experimental Methods

The cross-sectional morphology of the SnS thin films was characterized using SEM, as shown in Figure 1. The results confirm that the sample thicknesses are 82 nm, 199 nm, 616 nm, and 813 nm, respectively.

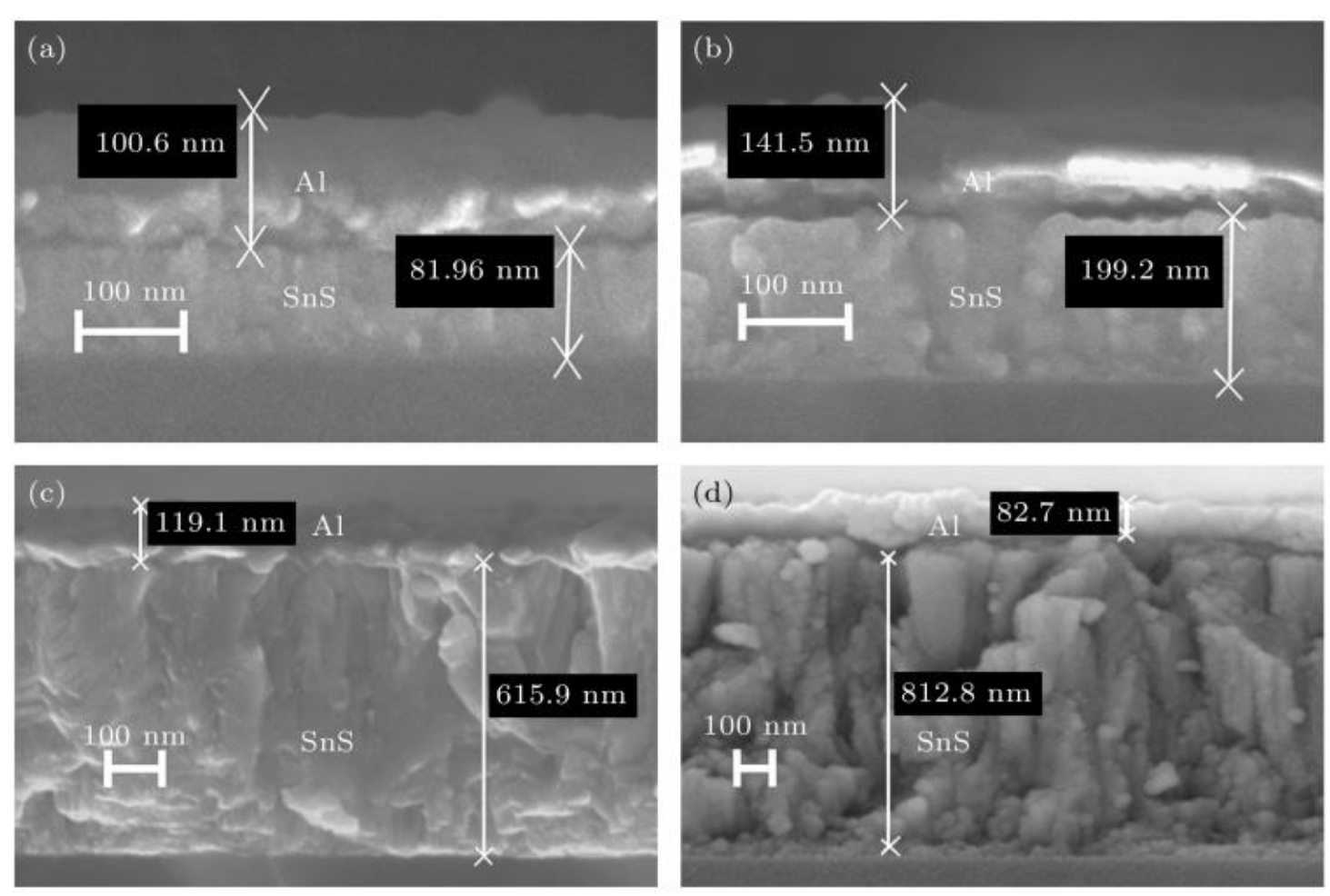

**Fig. 1 SEM cross-sectional images of SnS thin films with different thicknesses.**

To determine the elemental composition of the samples, this study characterized their elemental makeup using SEM and EDS. Figure 2 presents the EDS image of the 199 nm sample, indicating that the sample consists of Sn and S elements.

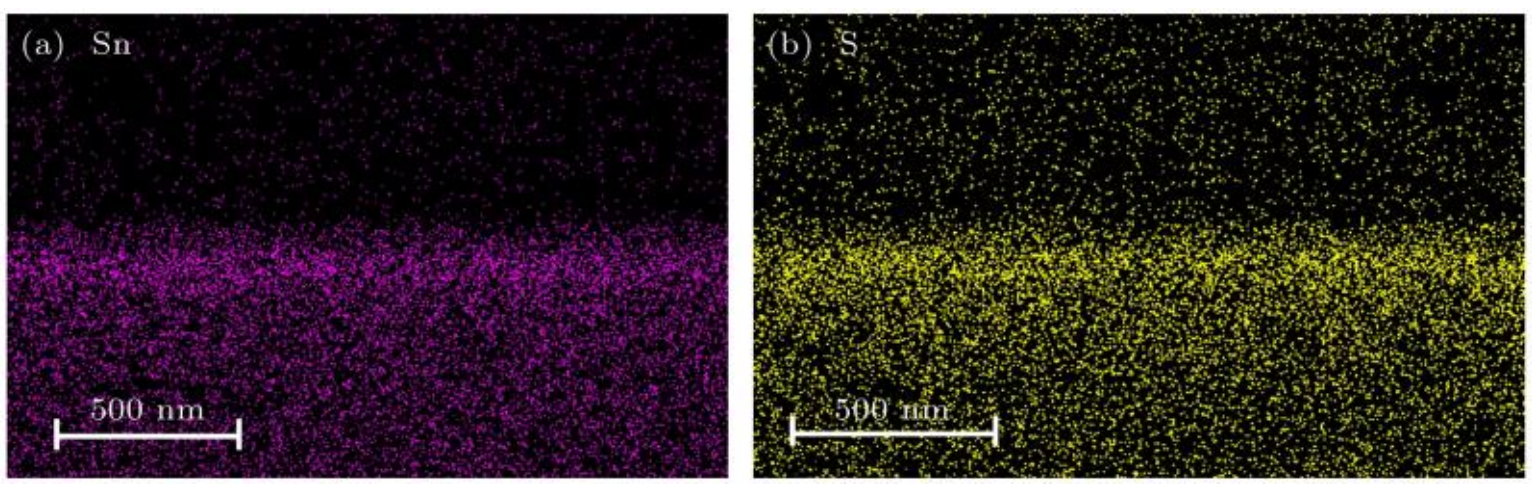

**Fig. 2 EDS elemental mapping of the 199 nm-thick SnS sample.**

The ZEM-3 is a high-precision integrated measurement instrument designed specifically for thermoelectric material research. Its core function lies in the simultaneous and accurate determination of key electrical transport and thermoelectric performance parameters, particularly electrical conductivity (or resistivity) and the Seebeck coefficient. The fundamental operating principle involves the synchronous measurement of the Seebeck coefficient and resistivity using the "static DC method" and the "DC four-probe method" within a precisely controlled temperature environment. The system positions the sample between two temperature-controlled probes. After the main heating furnace raises the sample to the set temperature, an integrated micro-heater applies a stable, small temperature difference ($\Delta T$). Four multifunctional probes, typically arranged in pairs, contact the side of the sample. One pair acts as a thermocouple to simultaneously collect the local temperature and the thermoelectric voltage ($\Delta V$) generated by the temperature difference. The Seebeck coefficient is then directly calculated based on the formula $S = -\Delta V/\Delta T$. For resistivity measurements, these probes serve as voltage sensors. A known constant current ($I$) is passed through the sample via outer electrodes, and the voltage drop ($V$) between two internal points is measured. The resistance is calculated according to Ohm's law $R = V/I$, and the resistivity ($\rho$) and electrical conductivity ($\sigma = 1/\rho$) are derived by incorporating the geometric dimensions of the sample. The entire measurement process is typically conducted in a low-pressure helium environment to ensure temperature uniformity and prevent sample oxidation. The system ensures high precision and reliability of data across a wide temperature range from room temperature to approximately 800 ℃ through automatic detection of ohmic contacts, elimination of parasitic thermoelectric potentials, and full computer-controlled programming.

Thermal conductivity is measured using the time-domain thermoreflectance (TDTR) method. Based on the photothermal reflectance principle, TDTR is a high-precision, high-resolution technique for measuring thermal properties. It is suitable for determining thermal transport properties such as the thermal conductivity and interfacial thermal conductance of nanoscale thin-film materials[25-27]. The basic principle of TDTR for measuring thermal physical properties is as follows: a pulsed laser beam is split into two beams. One beam heats the sample (pump beam), while the other monitors changes in the surface reflectivity of the sample (probe beam)[28]. When the temperature rise is small, the surface reflectivity exhibits a linear relationship with temperature, allowing the determination of surface temperature changes. These temperature changes correlate with the sample structure and thermal physical properties. The target thermal parameters are obtained by fitting the experimental data to a thermal diffusion model[29].

In the TDTR system, the overall system error can be derived using the propagation equation:

$$E_{\mathrm{r}} = \sqrt{\sum_{\xi} (S_{\xi}\sigma_{\xi})^2}, \tag{3}$$

Here, $E_{\mathrm{r}}$ denotes the error, $S$ represents the sensitivity, and $\xi$ refers to any input physical parameter that may affect the results. These parameters include the maximum uncertainty in the parameter settings for each sample layer, errors in spot size and temperature, and fitting errors caused by signal fluctuations.

# 3 Results and Discussion

Figure 3 illustrates the temperature dependence of the Seebeck coefficient for SnS films with different thicknesses. First, regarding the conduction type, the Seebeck coefficients of all samples remain positive throughout the tested temperature range. This indicates that the prepared SnS films exhibit p-type semiconductor characteristics[30,31]. Compared with undoped bulk SnS materials[32], the SnS films in this study demonstrate generally higher Seebeck coefficients. This enhancement is primarily attributed to the unique energy filtering effect in low-dimensional materials. Numerous grain boundaries and interfaces in the films form potential barriers. These barriers effectively scatter or filter out low-energy carriers, allowing only carriers with higher average energy to pass through, thereby increasing the Seebeck coefficient[33]. Furthermore, quantum confinement effects reshape the electronic band structure. The resulting density of states (DOS) peak near the Fermi level increases the energy derivative of carriers at the band edge, which further enhances the Seebeck coefficient[34].

The microstructure of the films, including grain size, grain boundary characteristics, and grain orientation, significantly influences the electronic density of states near the Fermi level[35]. In polycrystalline films, grain boundaries can serve as potential barriers to achieve the energy filtering effect[36]. Irregular atomic arrangements and local defects at grain boundaries may cause electronic band bending or form localized states. These changes alter the distribution of the density of states near the Fermi level[37]. Such localized states or barriers can induce quantum confinement effects similar to those in quantum dots. They introduce sharp features in the density of states near the Fermi level, thereby enhancing the Seebeck coefficient. Additionally, defects themselves can introduce extra electronic states. If these states are located near the Fermi level and exhibit sharp characteristics, they can also enhance the Seebeck coefficient[38].

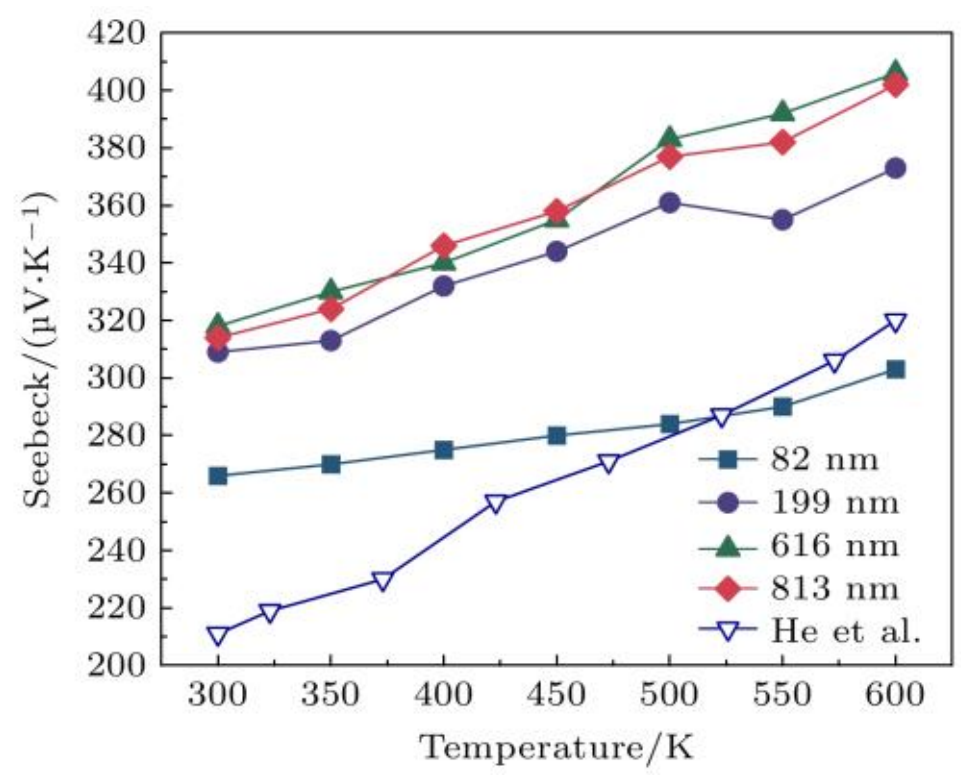

**Fig. 3 Temperature dependence of the Seebeck coefficient for SnS thin film samples with different thicknesses.**

The intrinsic link between electronic structure variations and Seebeck coefficient enhancement can be elucidated using Boltzmann transport theory and the Mott formula. The physical essence of the Seebeck coefficient is the entropy carried by charge carriers. More precisely, it reflects the energy asymmetry of transport properties near the Fermi level. When quantum confinement or energy filtering effects induce sharp features in the density of states near the Fermi level, such as resonance peaks or steep changes at band edges, the transport properties of electrons slightly above and below the Fermi level become distinctly different. High-energy carriers, such as those successfully traversing barriers in energy filtering, carry more thermal energy and participate more efficiently in charge transport. In contrast, low-energy carriers are suppressed. This selectivity for carrier energy causes the net transport current to be dominated by high-energy carriers. Consequently, the average entropy carried per carrier increases significantly, leading to an enhanced Seebeck coefficient. Furthermore, quantum confinement effects can enhance the Seebeck coefficient through band degeneracy. When multiple valleys converge in energy, the total electronic density of states increases substantially. This leads to an increased density-of-states effective mass, which is positively correlated with the Seebeck coefficient[39]. In summary, quantum confinement and energy filtering effects in polycrystalline thin films fundamentally enhance the Seebeck coefficient. They achieve this by introducing sharp density-of-states features near the Fermi level and selectively altering the average carrier energy. These effects do not exist in isolation but act together through complex coupling mechanisms[40].

Regarding the influence of thickness on thermoelectric performance, Figure 3

shows that the Seebeck coefficient of SnS thin films generally increases with thickness. This phenomenon can be explained by changes in carrier concentration. Typically, thinner films contain more lattice defects. These defects introduce additional electrons or holes, leading to higher carrier concentrations[41]. Since the Seebeck coefficient is generally negatively correlated with carrier concentration (Equation (4))[42,43], the improved crystallinity and reduced carrier concentration associated with increased thickness favor an enhanced Seebeck coefficient:

$$S = \frac{8\pi^2 m^* k_B^2 T}{3eh^2}\left(\frac{\pi}{3n}\right)^{2/3}, \quad (4)$$

In the equation, $m^*$ represents the density-of-states effective mass, $k_B$ denotes the Boltzmann constant, $h$ is the Planck constant, and $n$ indicates the carrier concentration.

As the temperature increases, the Seebeck coefficients of SnS films with varying thicknesses and bulk SnS[32] exhibit an upward trend. This behavior arises because grain boundaries reduce carrier mobility and influence the Seebeck coefficient through energy filtering[9]. However, the trend reverses in the high-temperature region. The rate of increase for the 82 nm sample slows down, while the 199 nm sample even shows a decline after 550 K. This suggests that carriers acquire sufficient kinetic energy at high temperatures to overcome grain boundary barriers, thereby weakening the filtering effect on low-energy carriers[9]. Additionally, the bipolar effect[44], which occurs readily in narrow-bandgap materials at high temperatures, contributes significantly to this decline. In this process, the excitation of minority carriers (electrons) offsets part of the Seebeck voltage. The 800 nm thick SnS film sample shows a relatively moderate growth rate in the low-to-medium temperature range. As the temperature rises further, the growth rate accelerates slightly, with no obvious peak or inflection point observed. This curve aligns closely with the 616 nm curve, and their values remain very similar. This indicates that thickness has a relatively weak impact on the Seebeck coefficient in these two thicker film samples. Their carrier transport mechanisms more closely resemble those of bulk materials. Ultimately, the 616 nm thick SnS film sample achieves the maximum Seebeck coefficient of 406 μV/K at 600 K.

Figure 4 illustrates the relationship between electrical conductivity and temperature for SnS films of different thicknesses and the corresponding bulk material[32]. Overall, the electrical conductivity of all film samples is lower than that of the bulk material. According to Equation (5) for electrical conductivity, this significant difference primarily stems from the microstructure of the thin films. The confined grain size and high density of lattice defects in the films cause strong carrier scattering. This results in carrier mobility that is far lower than that of the bulk material, thereby significantly reducing electrical conductivity[30]. Regarding temperature dependence, both the bulk material and SnS film samples of all thicknesses (82 nm, 199 nm, 616 nm, and 813 nm) exhibit typical metal-like conductive behavior[45] within the test range of 300—600 K. Specifically, electrical conductivity decreases as temperature rises. This occurs because increased temperature intensifies lattice vibrations and enhances phonon scattering. Phonons interact with carriers (electrons or holes) in the material, reducing the mean free path and lowering mobility[46-48]. When the rate of mobility decay exceeds the change in carrier concentration, the macroscopic result is a decrease in electrical conductivity. Specifically, the 616 nm thick SnS film is most sensitive to temperature changes. Its electrical conductivity drops substantially from $34.2\times10^3$ S/m to $11.8\times10^3$ S/m. Meanwhile, the 199 nm sample reaches its minimum electrical conductivity of $10.9\times10^3$ S/m at 600 K.

$$\sigma = ne\mu, \tag{5}$$

In the equation, $n$ denotes the carrier concentration, $\mu$ represents the carrier mobility, and $e$ is the elementary charge. Equation (5) indicates that both carrier concentration and carrier mobility are positively correlated with electrical conductivity.

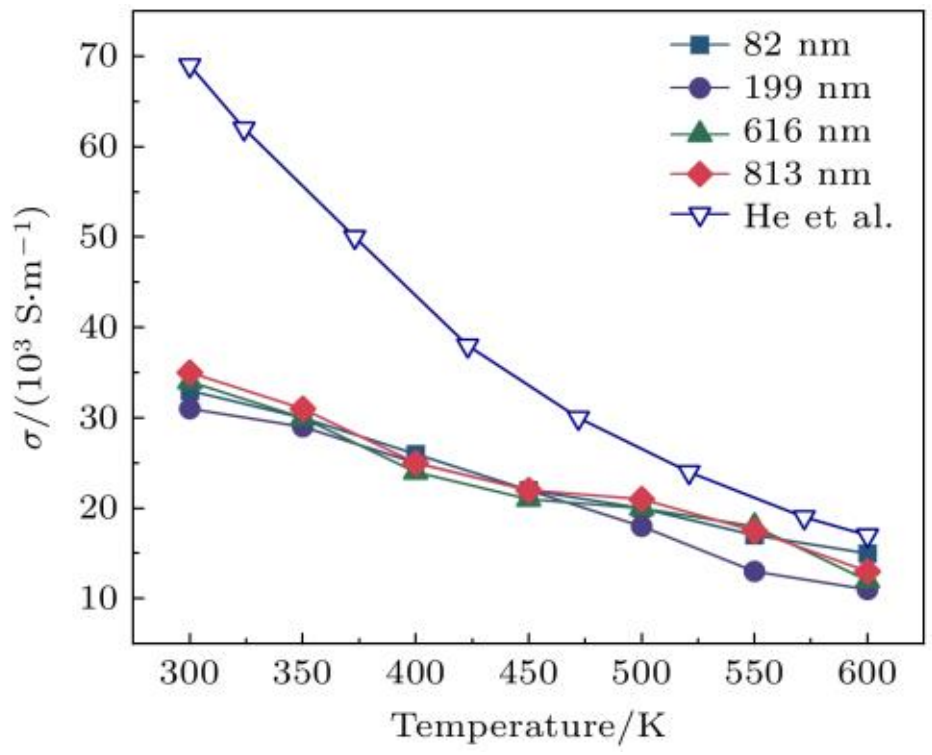

**Fig. 4 Temperature dependence of the electrical conductivity for SnS thin film samples with different thicknesses.**

Figure 5 illustrates the variation of the power factor (PF) with temperature for SnS films of different thicknesses and bulk material[32]. Overall, the PF of bulk SnS decreases monotonically as temperature rises. The film samples exhibit a similar downward trend: the PF values for the 82 nm and 199 nm SnS films drop to minimums of 13.8 $\mu W\cdot cm^{-1} K^{-2}$ and 15.2 $\mu W\cdot cm^{-1} K^{-2}$, respectively, at 600 K. Notably, although the 800 nm film generally shows a declining trend, it maintains excellent performance near 500 K, with a PF value as high as 29.8 $\mu W\cdot cm^{-1} K^{-2}$. The non-monotonic evolution of the PF with temperature, characterized by an initial decrease, followed by a recovery, and then a further decline, essentially results from the temperature-dependent competition between the Seebeck coefficient and electrical conductivity. In the low-temperature range, the rate of decay in electrical conductivity exceeds the growth rate of $S^2$, causing the PF to decrease. In the medium-temperature range, the increase in $S^2$ temporarily surpasses the decay in electrical conductivity, driving the PF to recover. In the high-temperature range, the accelerated decay of electrical conductivity becomes dominant again, ultimately leading to a continuous reduction in the PF.

This trend reflects the competitive mechanism between the Seebeck coefficient and electrical conductivity. Although electrical conductivity decreases with rising temperature, the enhancement of the Seebeck coefficient in samples of different thicknesses mitigates the reduction in PF to some extent. Most critically, as shown in Figure 5, the PF values of the 199 nm, 616 nm, and 813 nm SnS films are generally higher than those of the bulk material in the 300—600 K range. This indicates that the significant increase in the Seebeck coefficient achieved through low-dimensionalization successfully compensates for, or even exceeds, the loss in electrical conductivity caused by grain boundary scattering. This confirms that at the nanoscale, optimizing the band structure[49] and leveraging the energy filtering effect[50] can effectively break the trade-off between thermoelectric parameters, thereby yielding thermoelectric performance superior to that of bulk materials.

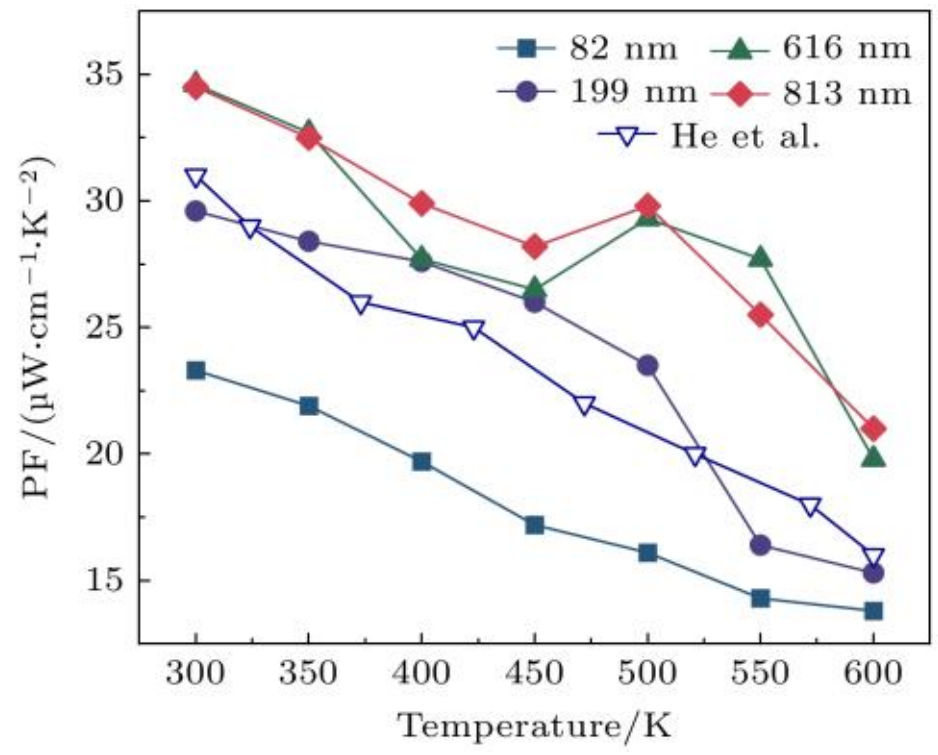

**Fig. 5 Temperature dependence of the power factor for SnS thin film samples with different thicknesses.**

Figure 6 presents the experimental thermal diffusion signals and theoretical fitting curves for SnS films of four different thicknesses within the temperature range of 300-600 K. The experimental data exhibit excellent agreement with the theoretical model, confirming the reliability of the adopted heat transport model and the extracted thermal conductivity values.

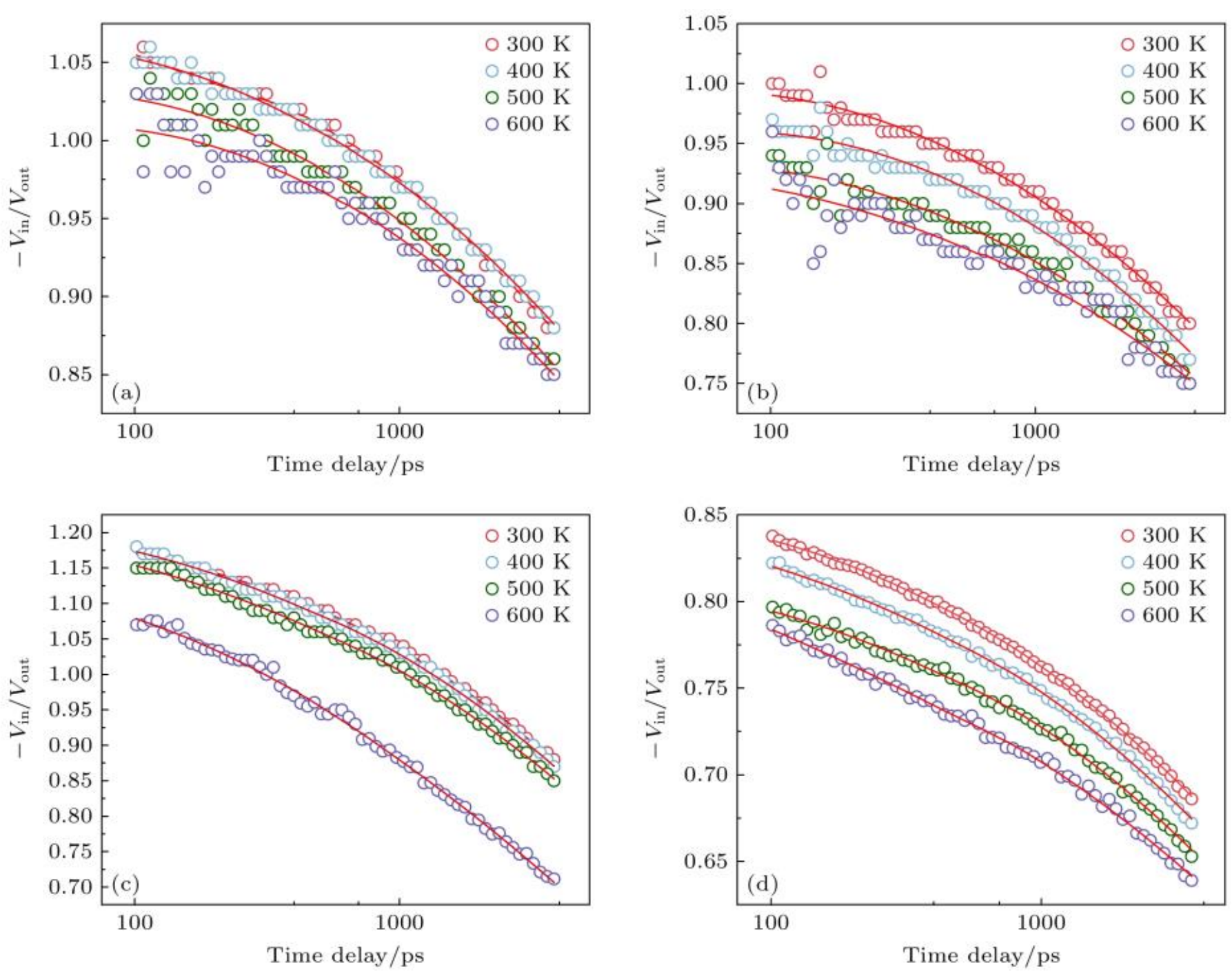

**Fig. 6 Experimental signals and fitting results for SnS films with four thicknesses at temperatures ranging from 300 to 600 K. Hollow circles represent experimental test data, while solid lines denote theoretical fitting curves: (a) 82 nm; (b) 199 nm; (c) 616 nm; (d) 813 nm**

From a microscopic perspective, the total thermal conductivity of the sample comprises electronic and phonon contributions. According to the Wiedemann-Franz law, electronic thermal conductivity is proportional to electrical conductivity. Since the electrical conductivity of the SnS films in this study is relatively low (as shown in Figure 4), the electronic contribution to total thermal conductivity is minimal[51] and exhibits a weak non-monotonic variation with thickness. Consequently, changes in the total thermal conductivity of the films are primarily dominated by phonon thermal conductivity.

Figure 7 illustrates the temperature dependence of thermal conductivity. The results indicate that the thermal conductivity of the films is significantly lower than the bulk value and decreases markedly with reduced film thickness. This phenomenon is mainly attributed to the classical phonon size effect: as thickness decreases, boundary scattering of phonons is enhanced due to the polycrystalline nature of the films[52]. This strong scattering effect significantly shortens the phonon mean free path, thereby substantially suppressing lattice thermal conductivity[53,54]. Notably, although thickness modulation significantly reduces thermal conductivity, electrical performance is not compromised to the same extent, as evidenced by the power factor results (Figure 5). This suggests that electrical and thermal transport can be partially decoupled at the nanoscale, offering new insights for designing high-performance thermoelectric materials.

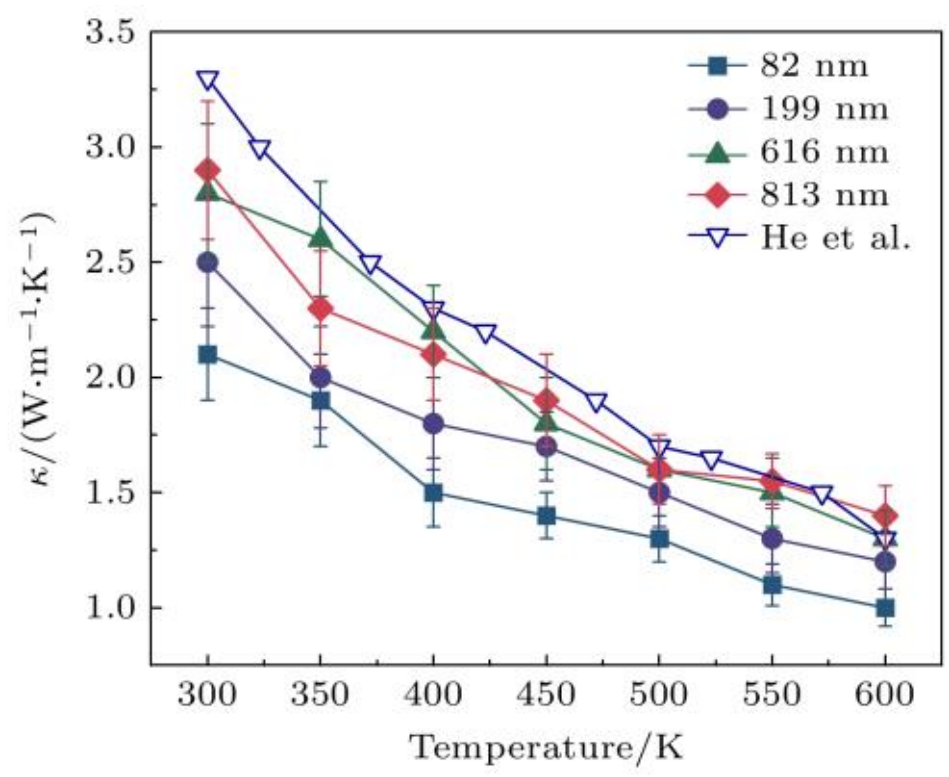

**Fig. 7 Temperature dependence of thermal conductivity for SnS thin-film samples with varying thicknesses.**

Furthermore, Figure 7 shows that the thermal conductivity of SnS thin films decreases with increasing temperature. This trend arises because elevated temperatures intensify phonon-phonon scattering, which effectively truncates the phonon mean free path and hinders lattice heat transport[55]. Notably, although interfaces and size effects in nanofilms complicate phonon scattering mechanisms, the enhancement of phonon-phonon scattering becomes more pronounced at high temperatures. Consequently, the material still follows the general rule that lattice thermal conductivity decays with temperature[15].

It is worth noting that the phonons contributing most to thermal conductivity in SnS have mean free paths primarily distributed in the 10—300 nm range, with peaks typically concentrated between 50 and 150 nm. Specifically, at room temperature, approximately 70%—80% of the thermal conductivity of SnS is contributed by phonons with mean free paths less than 200 nm. Among these, phonons with mean free paths in the 50—150 nm range make the most significant contribution, exhibiting the highest single-point contribution density[56]. In this work, the thickness of 82 nm is below the upper limit of the peak region for the mean free paths of the dominant contributing phonons in SnS, and is even lower than the characteristic mean free path of typical heat-carrying phonons. This implies that for 82 nm thick films, long-mean-free-path phonons, which contribute significantly to thermal conductivity, undergo strong boundary scattering[57,58]. When the characteristic dimension of a material, such as film thickness, is comparable to or smaller than the phonon mean free path, boundary scattering becomes the dominant mechanism, significantly reducing the thermal conductivity of the material[51]. This leads to a significant suppression of the transport of long-mean-free-path phonons, thereby explaining the marked decrease in thermal conductivity. In contrast, thicknesses of 616 nm and 813 nm far exceed the mean free paths of the vast majority of heat-carrying phonons in SnS. This means that for 616 nm and 813 nm thick films, phonons are more likely to undergo internal scattering, such as phonon-phonon scattering and defect scattering, before reaching the film boundaries, rather than experiencing boundary scattering[59]. Therefore, the limiting effect of boundaries on phonon transport is relatively weak, and their thermal conductivity approaches the limit of bulk materials.

In summary, the decrease in thermal conductivity of SnS thin films with decreasing thickness observed in this study is not a simple "size effect." Instead, it results from the complex interaction between the distribution of phonon mean free paths and the characteristic dimensions of the thin films. The 82 nm film thickness effectively truncates the portion of phonons that contribute most to thermal conductivity, leading to a significant drop in thermal conductivity. In contrast, the 616 nm and 813 nm film thicknesses are sufficiently large to allow most heat-carrying phonons to undergo diffusive transport, resulting in thermal conductivity values close to those of bulk materials.

Based on the measured experimental data for electrical and thermal transport, we calculated the thermoelectric figure of merit for SnS thin films, as shown in Figure 8. Although both the power factor and thermal conductivity decrease with increasing temperature, the more pronounced decline in thermal conductivity leads to an increase in the *ZT* value. This indicates that within the temperature range of this study, thermal properties exhibit stronger temperature sensitivity than electrical properties, thereby driving the overall upward trend in *ZT* values. As shown in Figure 8, the *ZT* values of all SnS thin film samples prepared in this study are significantly superior to those of traditional bulk materials. Notably, the 616 nm thin film sample achieved a *ZT* value of 1.01 at 550 K, representing an approximately 1.5-fold improvement over bulk materials. In the high-temperature range, the *ZT* values of the 616 nm and 813 nm SnS thin films remain at a high level of approximately 0.9, reflecting that films of these thicknesses maintain good thermoelectric performance at elevated temperatures. This performance enhancement confirms the effectiveness of the "low-dimensionalization" strategy. The physical mechanisms mainly include two aspects: in terms of electrical transport optimization, quantum confinement effects introduced by dimensionality reduction reshape the electronic band structure, effectively enhancing the Seebeck coefficient; in terms of thermal transport suppression, interfaces and boundaries at the nanoscale enhance phonon scattering, significantly reducing lattice thermal conductivity[55]. In conclusion, reducing material dimensionality has been proven to be an efficient approach to breaking through the bottleneck of SnS thermoelectric performance by synergistically optimizing electron and phonon transport properties.

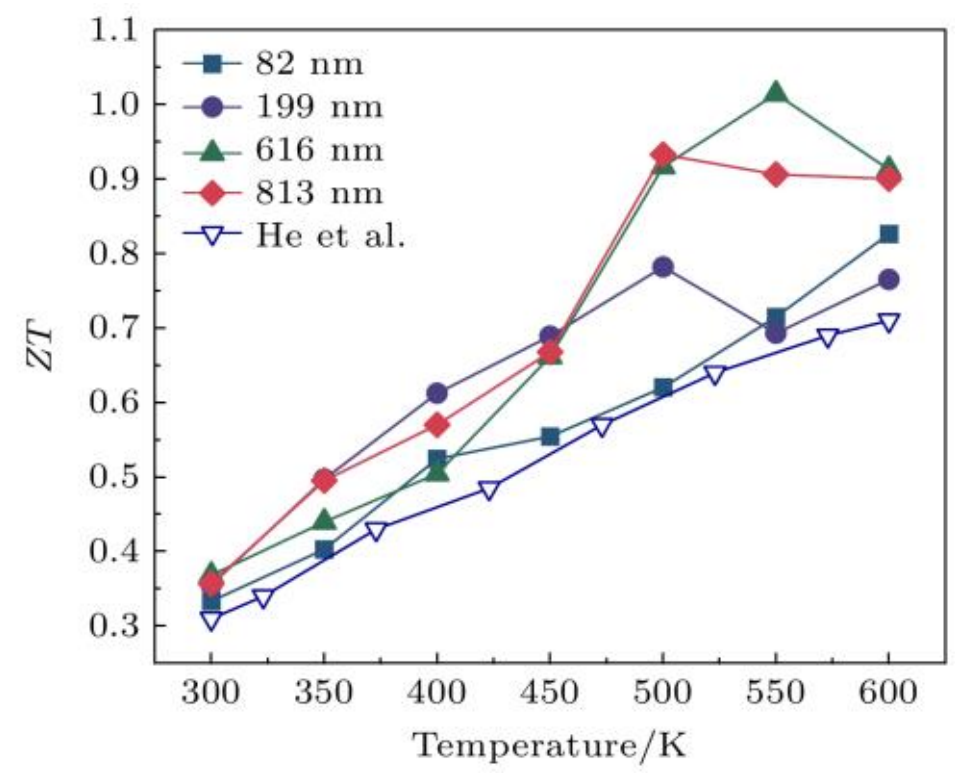

**Fig. 8 *ZT* values of SnS thin films in the temperature range of 300-600 K.**

# 4 Conclusion

This study comprehensively characterized the structural and thermoelectric properties of SnS thin films with four thicknesses (82 nm, 199 nm, 616 nm, and 813 nm). The results indicate that low-dimensionalization is an effective strategy for enhancing the thermoelectric performance of SnS. Furthermore, the figure of merit (*ZT*) exhibits significant dependence on both thickness and temperature. The main findings are as follows: 1) Compared with bulk materials, SnS thin films of all four thicknesses demonstrate superior thermoelectric performance. This enhancement primarily stems from quantum confinement effects, energy filtering effects, and enhanced phonon scattering induced by the low-dimensional structure. 2) As temperature increases, the Seebeck coefficient rises due to energy filtering, while electrical conductivity declines because of enhanced phonon scattering. Thermal conductivity also decreases due to intensified phonon-phonon scattering. The combined effect of these factors causes the *ZT* value to generally increase with rising temperature. This study deepens the understanding of thermoelectric transport mechanisms in SnS thin films. It provides important theoretical basis and experimental reference for optimizing material design and fabrication. Consequently, it offers significant scientific guidance for potential applications and further development in fields such as waste heat recovery and solid-state refrigeration.